# Modulating weak protein-protein cross-interactions by addition of free amino acids at millimolar concentrations

Pamina M. Winkler[1], Cécilia Siri[1], Johann Buczkowski[2], Juliana V. C. Silva[2], Lionel Bovetto[2], Christophe Schmitt[2], Francesco Stellacci[1]*

[1]*Laboratory of Supramolecular Nanomaterials and Interfaces, Ecole Polytechnique Fédérale de Lausanne (EPFL), Station 12, 1015 Lausanne, Switzerland.*
[2]*Nestle Research, Nestlé Institute of Food Sciences, Vers-chez-les-Blanc, CH-1000 Lausanne 26, Switzerland*

*KEYWORDS: long-range interaction, protein aggregation, protein stabilization, low molecular weight ligands, amino acids, lysozyme, whey protein isolates*

**ABSTRACT:** In this paper, we quantify weak protein-protein interactions in solution using Cross-Interaction Chromatography (CIC) and Surface Plasmon Resonance (SPR) and demonstrate that they can be modulated by the addition of free amino acids. With CIC, we determined the second osmotic virial cross-interaction coefficient ($B_{23}$) as a proxy for the interaction strength between two different proteins. We perform SPR experiments to establish the binding affinity between the same proteins. With CIC, we show that the amino acids proline, glutamine, and arginine render the protein cross-interactions more repulsive or equivalently less attractive. Specifically, we measured $B_{23}$ between lysozyme (Lys) and bovine serum albumin (BSA) and between Lys and protein isolates (whey and canola). We find that $B_{23}$ increases when amino acids are added to the solution even at millimolar concentrations, corresponding to protein: ligand stoichiometric ratios as low as 1:1. With SPR, we show that the binding affinity between proteins can change by one order of magnitude when 10 mM of glutamine are added. In the case of Lys and one whey protein isolate (WPI) it changes from the mM to the M, thus by three orders of magnitude. Interestingly, this efficient modulation of the protein cross-interactions does not alter the protein's secondary structure. The capacity of amino acids to modulate protein cross-interactions at mM concentrations is remarkable and may have an impact across fields in particular for specific applications in the food or pharmaceutical industries.

## INTRODUCTION

Weak protein (cross-) interactions are ubiquitous in nature and pivotal to many cell functions. We define weak protein-protein interactions as cross-interactions whose standard free energy does not exceed four times the thermal energy ($k_B T$). They are responsible for protein solubility but also for protein aggregation.[1,2] Recent studies suggest that small molecules acting as osmolytes are capable of screening these hydrophobic interactions and of rendering protein dispersions more stable.[3] This screening mechanism by osmolytes preventing aggregation has been observed in highly crowded environments such as the nucleus of the cell.[4–6] Yet, protein stability against aggregation is an important topic also beyond the context of cell biology, in fields where dilute solutions are prevalent. In the food industry, protein aggregation in dilute solutions plays a key role in a whole range of processes and phenomena. For instance, the precipitation of salivary proteins, including lysozyme and proline-rich proteins when interacting with certain food proteins, is assumed to be responsible for the perception of the astringent and dry mouthfeel sensation.[7] Among those food proteins leading to an astringency sensation when interacting with salivary proteins, whey and plant proteins are found to be majorly present in commercial protein isolates.[8,9]

Typically, the addition of salts at high concentrations to protein solutions enhances unfavorable interactions leading to a destabilization of the dispersions and to protein aggregation.[2,10] Other molecules can destabilize protein solutions because they favor denaturation (e.g. urea).[11,12] There has been intense research in the investigation of small molecules capable of stabilizing protein solutions. Trimethylamine N-oxide (TMAO) is considered the quintessential stabilizing molecule. It is believed that TMAO counteracts denaturing effects. Amino acids such as glycine and structurally equivalent ligands like betaine have been shown to be another class of small molecules that stabilizes proteins in aqueous solutions.[13]

In this paper, we show that amino acids have an effect on dilute protein solutions that is opposite to the typical effect of salts, *i.e.* that they reduce the net, unfavorable interactions between proteins. When studying dilute protein solutions, we used as a proxy for their stability two quantitative values, namely, the second osmotic virial cross-interaction coefficient ($B_{23}$) and the equilibrium dissociation constant ($K_D$).[14–17] We present here quantitative data showing that proline, glutamine, and arginine significantly change both values of $B_{23}$ and $K_D$ of selected protein couples even at relatively low concentrations. For example, we show that the $K_D$ of lysozyme to a-lactalbumin increases of one order of magnitude when 10 mM of glutamine is added to the solution.

### RESULTS and DISCUSSION

**Cross-interactions between lysozyme and other proteins are repulsive or attractive depending on the net charge of the proteins.**

To quantify the influence of small molecules on protein-protein cross-interactions, we applied Cross-Interaction Chromatography (CIC), an extension of Self-Interaction Chromatography (SIC) to quantify the cross-interaction between lysozyme (Lys) and a different protein of interest (refer to **Experimental Section** for detailed explanations).

As a proof-of-principle of CIC, as an extension of SIC, we quantified the interaction in terms of $B_{23}$ between Lys and bovine serum albumin (BSA) in 50 mM sodium phosphate buffer at pH ~6.9. In **Figure 1 A** we show a characteristic elution profile for Lys-BSA in comparison to the one of the self-interacting Lys in the same buffer conditions. The measured retention volume for Lys-BSA is (1.98 ± 0.04) ml which is eluted ~0.2 ml later than the Lys-Lys peak retention volume. With the determined retention volume at the peak position of the elution profile, we calculated the corresponding $B_{23}$ as explained in detail in the **Experimental Section**. For Lys-BSA, we obtained a $B_{23}$ value of $(-1.6 \pm 0.4) \cdot 10^{-4}$ mol·ml/g$^2$ indicating that the interaction is slightly attractive, given that the obtained $B_{23}$ value is negative. In contrast, the obtained data for the self-interacting lysozyme yielded a positive value for $B_{22}$ (= $3 \pm 0.2$) $10^{-4}$ mol·ml/g$^2$. At the buffer condition tested at pH~, Lys and BSA are oppositely charged. Thus, we expected a $B_{23}$ value for Lys-BSA hinting at attractive interactions. This is in agreement with our obtained data and also confirmed by the literature ones.[16,18] A brief remark on the peak shape of measured SIC/CIC elution profiles is that it contains several additional information that can be derived by a more elaborate peak deconvolution analysis. The peak width is a convolution of the molecular weight distribution with the interaction strength of the eluted protein with the grafted protein on the column. By visual inspection of the Lys-BSA peak shape to the one obtained for the Lys-Lys self-interaction (see **Figure 1 A**), it can be stated that the Lys-BSA peak is broader. This can be expected since BSA possesses a larger molecular weight, larger hydrodynamic radius and has a natural propensity to form dimers and trimers in buffer conditions containing salts. Lastly, note that we assume that the grafted Lys exhibits a random distribution. However, there still might be differences between bulk measurements and Lys grafted on the surface of a chromatography column. Overall, we have successfully managed to establish the CIC approach on the model system of the two globular proteins Lys-BSA and to quantify the cross-interaction strength in terms of $B_{23}$.

Next, we assessed the interaction strength of Lys with different commercial protein isolates. The protein isolates we tested are two types of whey protein isolates (WPIs), one enriched with b-lactoglobulin (BLG), referred to as WPI BLG, and the other one is enriched with a-lactalbumin (ALAC), referred to as WPI ALAC hereafter. In addition, a canola protein isolate rich in napin was tested and referred to as CPI NAP hereafter in our standard buffer condition. The isoelectric points have been found to be very close between the two WPIs, namely to be at pH~4.7 for WPI BLG and at pH~5.0 for WPI ALAC (**SI Figure 1 A and B, respectively**). From turbidity measurements it can be concluded that WPI BLG interacts significantly stronger with lysozyme than WPI ALAC (**SI Figure 2**).

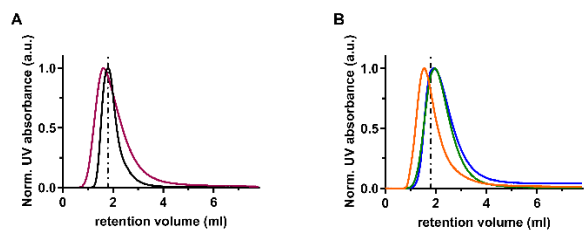

**Figure 1.:** Characteristic elution profiles measured by the CIC approach monitoring the interactions of Lys-BSA (purple) in comparison to the self-interacting Lys-Lys (black) **(A)** and of Lys-WPI BLG (blue), Lys-WPI ALAC (green), Lys-CPI NAP (orange) in comparison to the self-interacting Lys-Lys (dash-dotted black line) **(B)** in 50 mM sodium phosphate buffer at pH ~6.9.

As shown in **Figure 1 B,** the elution profiles corresponding to cross-interactions between Lys and the three different protein isolates studied are distinctly shifted in comparison to the peak position of the Lys-Lys self-interaction profile (black dotted line). These shifted peaks indicate that distinctly different interactions occur between the grafted Lys and the injected protein isolates. The obtained $B_{23}$ values are $(-1.0 \pm 0.3) \cdot 10^{-4}$ mol·ml /$g^2$ for Lys-WPI BLG and $(-0.7 \pm 0.4) \cdot 10^{-4}$ mol·ml /$g^2$ for Lys-WPI ALAC. Both $B_{23}$ values are negative revealing that the interactions between lysozyme and both WPIs are attractive. However, for the measured Lys-CPI NAP cross-interaction, we obtain a value of $B_{23} = (4.6 \pm 0.3) \cdot 10^{-4}$ mol·ml /$g^2$. This indicates that the interaction between Lys and CPI NAP is repulsive. Note that CPI NAP, a protein extracted from canola seeds, is considered to be the plant-based structural analogue of lysozyme.[19] They share similar molar mass, the secondary structure rich in helices, 4 disulfides bonds and a high isoelectric point which was measured to be at pH~7.2 (**see SI Figure 3**).Thus, at the studied pH of ~6.9 the two interacting proteins, Lys and CPI NAP, both possess a positive net charge which explains the observed repulsive nature of their interaction.

**The presence of proline at mM concentrations increases repulsive interactions between Lys and BSA.**

After having determined the $B_{23}$ values for protein cross-interactions under standard buffer conditions, we monitored the influence of small molecules, namely the addition of various amino acids to the buffer solution. In **Figure 2** and **Figure 3,** we report on the change of $B_{23}$ ($\Delta B_{23}$) to focus on solely detecting the impact of the added amino acid on the protein cross-interactions and removing all other effects (e.g. of buffers and salts, see the **Experimental Section** for more details). In **Figure 2,** we examined the influence of proline on the Lys-BSA interaction in terms of $\Delta B_{23}$. There is clearly a significant influence of proline that renders the cross-interaction between Lys and BSA more repulsive or equivalently less attractive ($\Delta B_{23} > 0$, **Figure 2**). In particular, it should be recognized that a noticeable effect of proline is already observed at an addition of 10 mM, the lowest concentration probed. The injected protein concentration of BSA is 20 mg/ml corresponding to 0.3 mM which translates to a minimal stoichiometric protein: ligand ratio of 33.

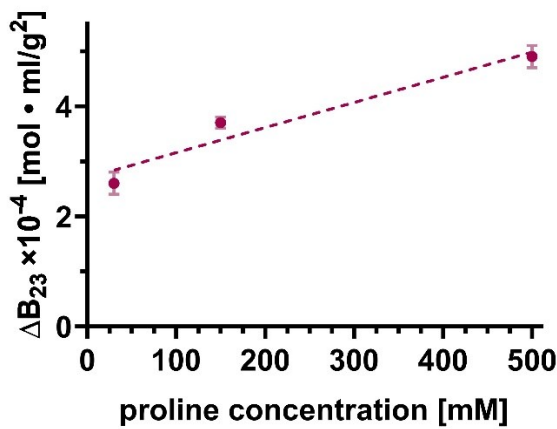

**Figure 2.:** Change of $B_{23}$ ($\Delta B_{23}$) for the interaction between lysozyme grafted to the column and BSA (at 20 mg/ml) in the presence of added proline dissolved in 50 mM sodium phosphate buffer at pH ~6.9. The error bars reflect the measurement uncertainties (std of $\Delta B_{23}$).

**The impact of various amino acids on the weakly interacting protein cross-interactions requires minimal concentrations.**

As observed for the Lys-BSA interaction, the effect of amino acids rendering the protein cross-interaction more repulsive holds true for the cross-interactions between Lys and different protein isolates as shown in **Figure 3**. Furthermore, the three amino acids tested appear to yield the same influence on the two interacting proteins regardless of their respective charge and/or polarity. Note that proline is a nonpolar amino acid, glutamine is polar but not charged and arginine is a basic, positively charged amino acid. The mentioned effect manifests itself as an increasingly positive $B_{23}$ value ($\Delta B_{23} > 0$) with increasing amino acid concentration in comparison to the one obtained in the buffer condition.

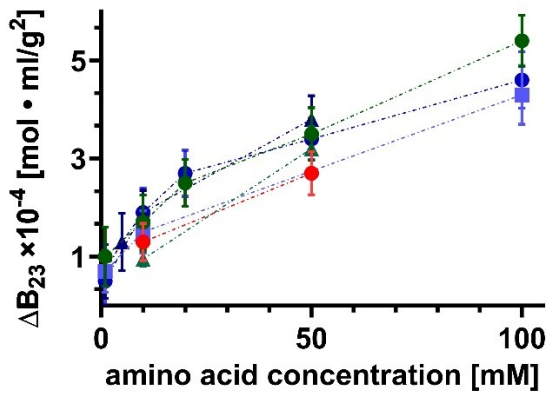

**Figure 3.:** $\Delta B_{23}$ for the interaction between lysozyme, grafted to the column, and the two whey protein isolates, WPI BLG (circles) and WPI ALAC (squares), and the canola protein isolate, CPI NAP (triangles), is displayed for different amino acids (arginine (blue), glutamine (green) and proline (red) ) at 1-100 mM dissolved in the protein injected and in the 50 mM sodium phosphate buffer at pH ~6.9. The error bars reflect the measurement uncertainties (std of $\Delta B_{23}$).

Strikingly, we observe this influence at added amino acid concentrations as low as at 1 mM for the two different WPIs interacting with Lys. This detectable positive shift of $B_{23}$ to more positive values measured at concentrations as minimal as 1 mM corresponds to a stoichiometric protein: ligand ratio of ~0.8 for WPI with a molecular weight slightly larger than Lys.

**The addition of mM concentrations of amino acids to Lys and WPI individually or to Lys interacting with WPI does not affect the secondary structure of the proteins.**

To verify if the presence of amino acids influences the secondary structure of the protein-protein cross-interactions, we performed Circular Dichroism (CD) experiments. The CD signal was monitored for the individual proteins as well as for mixtures interacting in the buffer first and then in the presence of amino acids at two different molar ratios as on display in **Figure 4** (refer to the **Experimental Section** for details).

In Figure 4 A-C we are comparing as a first visual inspection, the normalized CD curves for Lys (**Figure 4 A**) and WPI BLG (**Figure 4 B**) alone as well as for Lys-WPI BLG (**Figure 4 C**) in the buffer solution with the ones measured in the presence of amino acids at different concentrations. Regardless of whether we added proline at a concentration of 1 or 10 mM or in the presence of 10 mM glutamine, the obtained curves still overlap with the respective curves of the protein tested in buffer solution (**Figure 4 A-C**). This suggests that the secondary shape is unaffected by the presence of amino acids.

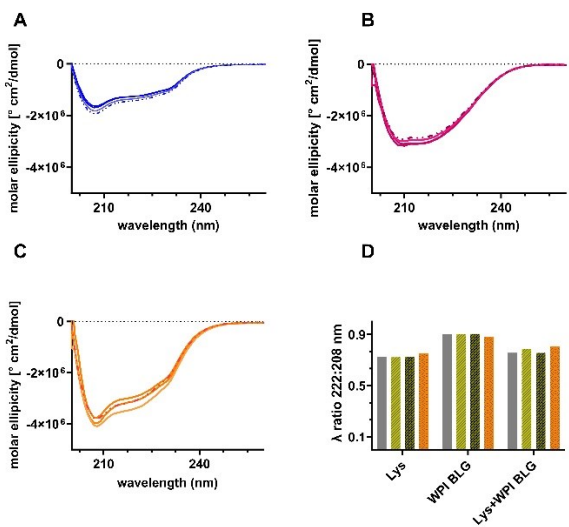

**Figure 4.:** Normalized CD curves for the individual proteins of lysozyme **(A)** and WPI BLG **(B)** at ~25 µM and for the interacting Lys-WPI BLG sample **(C)** at ~50 µM in buffer and in the presence of 1 mM proline and 10 mM proline and 10 mM glutamine. In **(D)** the calculated wavelength (λ) ratio between the CD signal measured at λ = 222 nm versus λ = 208 nm for the Lys and WPI BLG measured separately and together in buffer solutions (grey) and in the presence of 1 mM proline (light green) and 10 mM proline (dark green) and 10 mM glutamine (orange).

The calculated sum of the individual proteins versus the measured CD curves for proteins mixed together perfectly overlaps for both conditions, namely in absence and presence of added amino acids. As shown in **Figure 4D**, we calculated the wavelength ratio between the CD signal measured at 222 nm versus the one at 208 nm for the two proteins alone and measured together in buffer solutions and for the different amino acid conditions. Again, we see no significant change which supports our conclusion that the driving mechanism is not based on a molecular structural change of the weakly interacting proteins but is rather of an effective screening nature.

**The dissociation constant of Lys interacting with different proteins changes by orders of magnitude in the presence of millimolar concentrations of amino acids.**

To gain a deeper insight into the effect of amino acids on protein-protein cross-interactions with respect to their binding affinity, SPR experiments were carried out evaluating the equilibrium dissociation constant $K_D$. The interactions in the Lys-WPI BLG system are in the regime of weak, transient interactions. Accordingly, we expected the values for $K_D$ to be in the mM range, in contrast to antibody-antigen interactions which are tightly binding and possess $K_D$ values in the nM-pM range.[20] Weak interactions with mM binding affinities are rarely studied by SPR since the time resolution for resolving their fast association and disassociation kinetics is at the limit of the commercially available SPR instruments.[21] Here for this work, we optimized the concentration range of the injected protein to yield a reliable SPR response leading to a high degree of repeatability. Specifically, the data presented for the Lys-WPI BLG interaction was measured in over 30 individual measurements conducted on three different CM5 sensor chips and five different measurement days. A representative SPR sensorgram for Lys-WPI BLG in the concentration range of (0.1 – 10) mg/ml in PBS is shown in the **SI Figure 4 A**. From the sensorgram the maximum constant SPR response of the injected WPI BLG is computed and plotted against its respective concentration as shown in **Figure 5 A**. The fitting of this plot assuming a steady-state affinity model results in the characteristic $K_D$ value, a measure of the binding affinity, of the investigated protein cross-interaction. For the Lys-WPI BLG interaction in PBS we obtained a characteristic average $K_D$ value of (2 ± 1) mM.

Once we demonstrated a reproducible $K_D$ for the Lys-WPI interaction, we probe the influence of proline, glycine and arginine at 10 mM and 100 mM which was added to the WPI BLG injection as well as added to the running PBS buffer. A representative sensorgram for the Lys-WPI BLG interaction in the presence of arginine at 10 mM is shown in **SI Figure 4 B**. The corresponding steady state affinity fitting is displayed in **Figure 5 B**.

We observed a distinct influence on the $K_D$ being shifted from the mM to the M range in the presence of the amino acids. The large error is explained by the fact that the molar range pushes the experiments to the resolution limit of the instrument. Nevertheless, we consistently obtained a $K_D$ value for the Lys-WPI BLG interaction shifted by three orders of magnitude in the presence of any of the three amino acids tested.

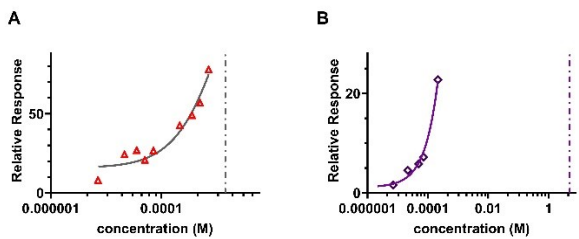

**Figure 5**. A representative steady state affinity fitting for the SPR response of WPI BLG interacting with lysozyme in PBS is shown in **(A)** yielding a $K_D$ value of (1.4 ± 0.2) mM which is indicated by the dashed line. In **(B)** the steady state affinity fitting is shown for a representative SPR response for WPI BLG interacting with lysozyme in the presence of 10 mM arginine yielding a $K_D$ of (5±1) M as indicated by the dashed line.

The question arises if the observed influence of amino acids on the binding affinity of protein-protein cross-interactions holds beyond the weakly binding interaction regime yielding mM values for $K_D$. As a first attempt, we performed SPR measurements to evaluate the binding affinities of Lys interacting with the protein a-lactalbumin (ALAC, molecular weight $M_W$ of ~14.2 kDa) and with the glycosaminoglycan hyaluronic acid (HA hereafter, low weight of $M_W$ ~100-500 kDa) in PBS buffer conditions and in the presence of 10 mM glutamine.

As shown in **Table 2** we obtained binding affinities of (8.9 ± 0.4)·$10^{-05}$ M for Lys-ALAC and of (1.4 ± 0.2)·$10^{-04}$ M for Lys-HA in buffer solution, respectively. The presence of 10 mM glutamine shifts both binding affinities by one order of magnitude. This indicates that the effect of amino acids is not unique for the mM range.

**Table 2.** Averaged $K_D$ values for the Lys-ALAC and Lys-HA cross-interactions in PBS buffer and in the presence of 10 mM glutamine measured by SPR experiments.

| Av. $K_D$ ± Std [M] | ALACc | HA |
|---|---|---|
| in PBS | (8.9 ± 0.4)·$10^{-05}$ | (1.4 ± 0.2)·$10^{-04}$ |
| 10 mM Gln | (6.2 ± 0.3)·$10^{-04}$ | (2.7 ± 0.9)$10^{-03}$ |

## CONCLUSIONS

In this work we quantified the effect of free amino acids on the interaction between lysozyme and a series of proteins in terms of changes in the second osmotic virial cross-interaction coefficient ($B_{23}$) and in the equilibrium dissociation constant ($K_D$). The protein combinations chosen were such that we had attractive (for Lys-BSA, Lys -WPI BLG and Lys-WPI ALAC) or repulsive interaction (Lys-CPI NAP) in 50 mM of phosphate buffer solutions at pH~6.9.

Regardless of the starting interaction regime for the specific protein system, we observed a significant influence of the added amino acid on all tested protein systems. In all cases, $B_{23}$ became larger and more positive indicating a net change in interaction towards a more repulsive regime. The effects we found are not small, i.e. for Lys-BSA interacting in presence of 500 mM added proline, we observed a value of $B_{23}$ that is more than double the initial one. In the case of lysozyme interacting with BSA or with any of the two whey protein isolates the initial $B_{23}$ value is negative, indicating a net attractive interaction. For BSA already 10 mM of added proline changes the sign of the $B_{23}$ value. For the whey protein isolates already 1 mM of added amino acid shifts the sign, however, note that the initial $B_{23}$ value is less negative. These data are significant since it shows that at protein to amino acid stoichiometric values as low as 0.8 we already observed a significant change in the interaction between proteins. A change in $B_{23}$ implies a change in the chemical potential of the two proteins, that in turn implies that the interaction between the protein has to fundamentally change.[22] A direct consequence of such protein cross-interaction is an (observable) change of its corresponding equilibrium dissociation constant $K_D$, a proxy for its binding affinity as well. We verified this statement by measuring changes in $K_D$ as a result of the addition of amino acids. The affinity measurements by SPR showed that 10 mM of added amino acid affects the binding constants expressed as a shift to weaker binding affinities. For the Lys-WPI BLG cross-interaction we report on a shift of $K_D$ by three orders of magnitude from the weakly interacting millimolar regime. In short, we show that free amino acids can modulate weak protein-protein cross-interactions as was previously reported for strongly binding immunoglobulins in presence of 5 mM histidine addition.[23]

We believe that the observed modulation of the protein cross-interaction by minimal concentrations of amino acids cannot be simply explained as an hydrotropic effect.[4,24] In fact, we do not see any threshold in this behavior and we observe the effect at low stoichiometries of protein: ligand (e.g. ~0.8) that can hardly be attributed to hydrotropic effects.

By means of CD measurements, we observed no changes on the secondary structure of the interacting proteins in the presence of amino acids. This implies that the modulating amino acids are only weakly interacting with the proteins. Also, that implies that the effect we have presented in this work, does not depend on protein conformational changes but it is the direct effect in the changes of the colloidal interaction between proteins. We believe that an amino acid can be present on the surface of a protein in a time-averaged fractional way, thus screening part of the protein interaction with other proteins (and with the solvent molecules) which leads to the change of its chemical potential. This small molecule effect already happens at low (e.g. millimolar) concentrations and can thus act as a powerful stabilizer for protein dispersions.

In this work, we show that this effect is rather broad. We investigated a range of protein systems, namely of two different WPI and of one CPI, which go beyond the most reported model proteins (i.e., Lys and BSA). In all cases the observed effects are comparable. Consistently with the proposed theory, the amino acid effects that we probed here are all similar in magnitude. This is because the root cause is a weak screening interaction of the amino acids with the proteins, that is not expected to change significantly across amino acids.

It is important to notice that the stabilization effects that we show here lead to large changes in the equilibrium dissociation constant. The mere presence of 10 mM glutamine in strongly binding systems such as Lys-HA ($K_D$ = 14 mM) and Lys - ALAC ($K_D$ = 0.9 µM) leads to a change in $K_D$ of close to one order of magnitude. This indicates that the effect of amino acids is not unique to weakly interacting protein systems. Furthermore, this highlights that free amino acids are potentially capable of modulating protein interactions across interaction regimes.

In the pharmaceutical field, amino acids such as arginine are employed as stabilizers in drug formulations without necessarily understanding the underlying mechanism. Here we have shown that the proteinogenic amino acids proline, arginine and glutamine have a fundamental stabilizing effect on protein-protein cross-interactions. This effect leads to significant alterations in the equilibrium dissociation constant. We foresee implications of this work on the use of amino acids in the design and preparation of protein-based liquid formulations, in both food and pharmaceutical industries. However, we still need to understand the extent of the capacity and the generality of free amino acids to modulate protein cross-interactions from future experiments.

## EXPERIMENTAL SECTION

### Materials

Hen egg-white lysozyme (14.3 kDa, ≥ 95%, purchased from Roche) was used as the primary model protein, and was stored at 4°C. For the grafting of the SIC column, a Tricorn 5/50 column (Cytiva, Column Volume of 1.178 ml) was manually grafted with lysozyme using as a resin TOYOPEARL®AF Formyl-650M chromatography particles, sodium cyanoborohydride, potassium phosphate and ethanolamine. The standard buffer used throughout the experiments was 50 mM sodium phosphate buffer at pH ≈ 6.9 consisting of monobasic and dibasic sodium phosphate in MilliQ water. Note that we chose this buffer as our standard one since it mimics the salivary conditions based on the average salt composition of saliva.[25,26] The commercial whey protein isolates were purchased from Agropur (Eden Prairie, MN, USA, reference BiPRO Alpha 9000 and BiPRO 9500) while the canola protein isolate was from Merit Functional Foods Inc. (MA, Canada, reference Puratein HS). Before the experiments the required proteins and protein isolates were weighed, typically at a protein content of 20 mg/ml, dissolved in the buffer solution and vortexed several times to obtain a homogeneous solution. All amino acids were purchased from Thermo Scientific being L-arginine, L-glutamine, L-proline, L-serine and L-glycine dissolved in the standard buffer.

### Column grafting

The experimental procedure for the custom-made column grafting with lysozyme for the SIC/CIC experiments in this study is a modified version of the protocol reported by Le Brun et al.[27] First, 3 ml of TOYOPEARL® AF-Formyl 650M particles were washed 5-7 times with our standard 50 mM sodium phosphate buffer. The recovered particles were mixed with lysozyme at 10 mg/ml dissolved in buffer solution. Then, 90 mg of sodium cyanoborohydride was added serving as an activator to covalently bind the particles and proteins. This solution was placed on a shaker for 12-14 hours overnight at room temperature. The next morning, the resin + lysozyme solution was washed with 200 ml of buffer using a membrane filter. A small sample was saved from the washing mixture to measure its lysozyme concentration indicating how much protein was washed away. Next to cap the residual matrix reactive groups, 20 ml of buffer with 1 M ethanolamine (1.22 g) and 20 mg sodium cyanoborohydride were prepared. The recovered resin particles grafted with lysozyme were added and the solution was placed on the shaker for 4 h at room temperature. To remove any unbound material, the particles were washed again with 200 ml of buffer from which was kept again a small sample. As a last step, the recovered grafted resin particles were dissolved in buffer solution in a graduated cylinder and were let to settle. To verify the amount of lysozyme bound as grafted protein to the resin particles, absorbance measurements using the Nanodrop instrument were performed on the three kept small samples of the stock solution, the solution after the first wash and after the second wash. All experiments were performed on SIC columns with a grafted surface coverage of lysozyme of ≈ 45%. To pack the Tricorn® 5/50 column the chromatography particles settled down in the graduated cylinder were prepared as a 58-60% slurry in buffer. The column was packed under pressure with the following flow rates: 0.75 ml/min for 15 minutes, then at 3 ml/min at 15 min and again at 0.75 ml/min for 30 min. At the end of the column packing as well as at each start of a measurement series, a column performance test was performed with 50 µL of a 20 vol% acetone solution to evaluate the reference elution volume. The column was stored at 4°C overnight and between experiment days.

### SIC/CIC experiments

To prepare a chromatography column for a CIC experiment, Lys is manually grafted on the column, as outlined in the previous section on **Column grafting,** and for the optimized signal-to-noise ratio of the elution profile we determined that the protein concentration should be in the range of ~20 mg/ml which was injected in all CIC experiments shown in this work.

SIC/CIC experiments were conducted to probe the lysozyme-lysozyme or the interaction between lysozyme and different proteins (BSA, WPI BLG, WPI ALAC and CPI NAP) in different solution environments (buffer alone or in the presence of an amino acid at different concentrations dissolved in buffer) from which the respective $B_{22}/B_{23}$ values were calculated. Before each measurement series, a column performance test was run with 20% Acetone in MilliQ water. For each run of the experiments, 50 µL of lysozyme at 20 mg/ml was injected. Samples were injected after 10× Column Volume and with a constant flow rate of 0.75ml/min at room temperature. The amino acids tested were proline, glycine, arginine, serine and glutamine and their concentration was varied between 5 mM and 1.2 M (considering their respective solubility limit). For each amino acid the concentration range was individually refined to be the lowest concentration possible (≥ 1mM) to observe a change in the $B_{22}/B_{23}$ value. The upper range limit was set by the fact that towards the solubility limit of an amino acid the buffer solution becomes turbid which very likely clogs and thus breaks the column. Therefore, to protect the grafted column the upper limit was chosen with caution in a range well below the solubility limit of the studied amino acid. In the case of glutamine, the least soluble amino acid tested here, given that the solubility limit is ~280 mM in aqueous buffer solution, the measurement range never surpassed 100 mM.

**Determination of $B_{22}/B_{23}$ by SIC/CIC**

When quantifying the interaction of two proteins in solution, a common approach is to use the virial equation of state.[17] In such equation, $B_{23}$ is the second order coefficient of the cross-term that depends on the concentration of protein 2 multiplied by the concentration of protein 3. This derivation uses statistical mechanics to link thermodynamics to the properties of molecules or proteins. The expansion of the osmotic pressure into its second virial coefficients allows for a measure of the non-ideality of the solution and has been used to quantify intermolecular forces between molecules in dilute solutions. In other terms, $B_{23}$ can be described by the cross-interaction energy or potential of mean force for two molecules as a function of separation distance and angular conformation.[16,17,28] A negative/positive $B_{23}$ is a signature for attractive/repulsive interactions, respectively, and when referenced to $B_{23}= 0$ that indicates an "ideal solution". The self-interaction second osmotic virial coefficient ($B_{22}$) is generally measured using traditional colloidal characterization techniques including static light scattering and Sedimentation Equilibrium Analytical Ultra-Centrifugation (SE-AUC) and Self-Interaction Chromatography (SIC).[22,28,29] SIC has the unique capability to be extended to Cross-Interaction Chromatography (CIC) quantifying the cross-interaction between two different proteins in terms of $B_{23}$.

In SIC/CIC experiments, one evaluates the interactions between the injected protein in the mobile phase and the immobilized protein grafted on the column in terms of a measured retention volume. To experimentally determine $B_{23}$ one will first compute the retention factor $k'$ from the measurement as follows:

$$k' = \frac{V_0 - V_r}{V_0}$$

where $V_0$ is the retention volume of non-interacting species which is calculated before each experiment with the column performance test using 20% acetone in MilliQ water and $V_r$ is the volume required to elute the injected protein in the mobile phase through the grafted column.

Then, $B_{23}$ [mol ml g$^{-2}$] can be computed as:

$$B_{23} = B_{HS} - \frac{k'}{\rho_s \cdot \Phi}$$

where $B_{HS}$ is the excluded volume or hard sphere contribution of the two interacting proteins, $\rho_s$ being the immobilization density, i.e. the number of covalently immobilized protein molecules per unit area of the bare chromatography particles and $\Phi = \frac{A_s}{V_0}$ is the phase ratio (i.e. the total surface available to the mobile phase protein. $B_{HS}$ is calculated as follows assuming a spherical shape:

$$B_{HS} = \frac{2}{3}\pi \cdot [r_2 + r_3]^3 \cdot \frac{N_A}{M_{2,3}^2}$$

where $r_{2,3}$ are the protein radii of the two proteins, $N_A$ refers to Avogadro's number and $M_2$ to the averaged protein molecular weight of both proteins. Lysozyme has a molecular weight of ≈ 14300 g/mol and a protein radius of (1.89 ± 0.03) nm.[30] For BSA the reported values for the radius is 3.48 nm and for the weight 66463 g/mol.[31]

The assumption here is that we are only measuring two-body interactions, i.e. one injected free protein interacts with only one immobilized protein molecule at a time. This is valid since protein-protein (cross)-interactions are dominantly of short-range nature, meaning that they are dominant over a smaller distance than the diameter of the proteins involved. This constraint can be guaranteed by controlling the immobilized proteins grafted onto an effectively flat surface being the column. The last assumption is that the injected free proteins interact only with immobilized proteins grafted onto the column and not with each other. This can be verified by determining the variation for the calculated $B_{22}$ value measured at a concentration of 5-30 mg/ml of the injected protein, here for lysozyme. For the Lys-Lys self-interaction the obtained $B_{22}$ value should remain constant. We determined this variation to be ~0.2 10$^{-4}$ mol ml g$^{-2}$ for Lys-Lys and considered this variation in our error analysis.

In the main text and figures we chose to report on the change of $B_{23}$ ($\Delta B_{23}$) being the absolute difference between the second osmotic virial cross-coefficient value in the presence of the amino acid and the one in the standard buffer. Hereby, variations between sample runs and measurement days due to differences of column grafting and other instrument variations are eliminated.

**SPR experiments and analysis**

SPR experiments were conducted on a Biacore 8K (Cytiva) at room temperature. We used the commercial CM5 Sensor S chip (Biacore) which is already functionalized with a carboxylated dextran matrix. To immobilize lysozyme on the chip, the standard routine for amino coupling was applied. This routine consists of an activation step with EDC and NHS, then lysozyme is linked to the matrix at a concentration of ~100 µg/ml in acetate buffer of pH 4.5.

This optimal pH was evaluated by a pH scouting step beforehand. After the lysozyme was coupled to the chip, the surface was passivated by ethanolamine. To determine the binding affinity we set up multi-cycle kinetic studies of WPI in PBS first and then in the presence of amino acids. To enhance the SPR signal and reduce the so-called gradient effect inducing a significant refractive index mismatch, we had to increase the salt content. This is the reason why we switched to PBS as the standard buffer for our SPR experiments. Note that only the ionic strength is higher but the pH remains in the neutral range. The use of PBS improved the signal-to-noise ratio of the SPR signal, but did not alter the obtained $K_D$ values. We determined that the optimized range for WPI BLG weakly and transiently interacting with the immobilized Lys onto the chip is to probe in the concentration range of 0.1 – 10 mg/ml of WPI BLG. We averaged over 10 experiments, each on > 4 channels with different degrees of Lys covalently immobilized on two separate chips. We applied the same concentration range for HA and α-lac. $K_D$ yields a measure of the binding affinity of the interaction.[15] The smaller the $K_D$ value, the higher the affinity between the two proteins. To determine $K_D$ from the SPR measurements, we used the Biacore™ Insight Software. From the measured SPR sensorgrams (which is SPR signal in Response Units (RU) vs time), we obtained the steady state binding levels ($R_{eq}$) vs analyte concentration ($C$) first and then by a fitting applying the steady state affinity model based on a 1:1 binding we determined $K_D$:

$$R_{eq} = \frac{C \cdot R_{max}}{K_D + C} + offset$$

With $R_{max}$ being the analyte binding capacity of the functionalized surface and the $offset$ the response at zero analyte concentration.

### CD experiments and analysis

All circular dichroism data were obtained on a Chirascan CD spectrometer (Applied Photophysics) using a quartz cuvette with a path length of 0.5 mm, allowing for a protein concentration of tens of µM. Lysozyme and WPI BLG were dissolved in our standard 50 mM sodium phosphate buffer at pH~6.9 to a final concentration of 25-50 µM and to this sample the amino acids were added. We monitored the influence of the added amino acid on the secondary shape at protein: ligand molar ratios of 1:2 and 1:20, which are similar to the CIC experimental conditions. The spectra were collected in the wavelength range of 190-280 nm with a scanning speed of 20 nm/min at room temperature. Each CD spectra is an average of three repeated measurements and is corrected for buffer and added amino acid absorption. Note that we always added the amino acids to the protein injection sample as well as that we dissolved the respective amino acid in the standard elution buffer to eliminate a gradient effect. We analyzed and visualized the spectra by ourselves in Excel and Prism 9 (GraphPad).

## AUTHOR INFORMATION

**Corresponding Author**
*E-mail: francesco.stellacci@epfl.ch

## Author Contributions
The manuscript was written through contributions of all authors. All authors have given approval to the final version of the manuscript.

## Funding Sources
The research of P.M.W. was funded by Société des Produits Nestlé S.A.

## ACKNOWLEDGMENTS
The authors gratefully acknowledge support and insightful discussions with Dr. Christoph Hartmann (Nestlé Research) and Dr. Kelvin Lau (EPFL) and Dr. Quy Ong (EPFL).